\documentclass[12pt]{article}

 \usepackage{amsmath,amsthm,amssymb}
 \usepackage{makeidx,epsfig,lscape}
 \usepackage{color,colortbl}
 \usepackage{fancyhdr}
 \usepackage{xcolor,pict2e}
\usepackage{latexsym}
\usepackage{graphicx,graphics}
\DeclareGraphicsExtensions{.bmp,.jpg,.eps,.pdf}
\usepackage{float,verbatim,url}
\usepackage{subfigure}

 \renewcommand{\footrulewidth}{0.5pt}

 \definecolor{myaqua}{rgb}{0.0,0.5,0.55}
 \definecolor{lightaqua}{rgb}{0.75,0.95,0.95}

 \usepackage[colorlinks = true,
            linkcolor = myaqua,
            urlcolor  = blue,
            citecolor = myaqua]{hyperref}

\usepackage{caption}

\def\lin#1#2{\textcolor[rgb]{0.6,0.6,0.6}{\vspace*{#1mm} \hrule
   height 3 pt \vspace*{#2mm}}}
\def\bt{\begin{tabular}}
\def\et{\end{tabular}}
\def\and{\mbox{ and }}

\def\1{{\bf 1}}

 \def\sectionn#1{\refstepcounter{section}{\color{myaqua}

 \vskip 6mm

 \noindent\Large\bf\thesection. #1}

 \vskip 3mm}

\begin{document}


 $\mbox{ }$

 \vskip 12mm

{ 

{\noindent{\huge\bf\color{myaqua}
  Inverse problem for time-series valued  \\[0.1in] computer model via scalarization}}
%
\\[10mm]
{\large\bf Pritam Ranjan$^1$, Mark Thomas$^2$, Holger Teismann$^2$, Sujay Mukhoti$^1$}}
\\[2mm]
{ 
 $^1$OM \& QT Area, Indian Institute of Management Indore, Indore, MP, India\\
\\[1mm]
$^2$Dept. of Mathematics \& Statistics, Acadia University, Wolfville, NS, Canada\\
Email:
\href{mailto:pritamr@iimidr.ac.in}{\color{blue}{\underline{\smash{pritamr@iimidr.ac.in}}}},

\lin{5}{7}

 { 
 {\noindent{\large\bf\color{myaqua} Abstract}{\bf \\[3mm]
 \textup{For an expensive to evaluate computer simulator, even the estimate of the overall surface can be a challenging problem. In this paper, we focus on the estimation of the inverse solution, i.e., to find the set(s) of input combinations of the simulator that generates (or gives good approximation of) a pre-determined simulator output. Ranjan et al. (2008) proposed an expected improvement criterion  under a sequential design framework for the inverse problem with a scalar valued simulator. In this paper, we focus on the inverse problem for a time-series valued simulator. We have used a few simulated and two real examples for performance comparison. }}}
 \\[4mm]
 {\noindent{\large\bf\color{myaqua} Keywords}{\bf \\[3mm]
 Calibration; Computer experiments; Contour estimation; Gaussian process model; Non-stationary process; Sequential design.
}


\lin{3}{1}

\sectionn{Introduction}

{ \fontfamily{times}\selectfont
 \noindent Experimentation with computer simulators have gained much popularity in the last two-three decades for applications where actual physical experiment is either too expensive, time consuming, or even infeasible. The applications range from drug discovery, medicine, agriculture, industrial experiments, engineering, manufacturing, nuclear research, climatology, astronomy, green energy to business and social behavioural research. A computer simulator (or computer model), built with the help of an application area expert, is often a mathematical model implemented in C/C++/Java/etc. which aim to mimic the underlying true physical phenomenon. Thus, the ultimate objective of the (unobservable) experiment with the true physical process (or phenomenon) can be fulfilled via the computer simulators.
\renewcommand{\footrulewidth}{0pt}

%

In this paper, we are interested in the inverse problem for deterministic simulators, i.e., find the input(s) of the simulator that corresponds to a pre-specified output. That is, if $g(x)$ represent the simulator response for an input $x$, then the objective is to find $S(g_0) = \{x\ :\ g(x)=g_0\}$.  Ranjan et al. (2008) proposed a sequential design framework for efficient estimation of the inverse solution when the simulator returns a scalar response, whereas the simulator under consideration in this paper gives time-series response. That is, we are interested in a \emph{functional inverse problem}. 

This research is motivated by two real-life applications. The first application comes from the apple farming industry in the Annapolis valley, Nova Scotia, Canada, where the objective is to find a suitable set of  parameters of the two-delay blowfly (TDB) model that corresponds to the reality. TDB model simulates population growth of European red mites which infest on apple leaves and diminish the quality of crop (Tiesmann et al. 2009). The data collection for the true population growth of these mites is very expensive, as the field expert would have to periodically count the mites on the leaves of apple trees in multiple orchards.  

The second application focuses on the calibration of a simulator which projects the inflation rate of a country over a period of time. Inflation or increase in overall price level is a key metric in determining the economic and financial health of a country, and the central banks are the key policy makers involved in such projections or setting up a target (\url{http://www.imf.org/external/pubs/ft/fandd/basics/target.htm}). To steer the actual inflation towards the target, the central banks control its driver, the interbank interest rate.  We use a computer model called Chair-The-Fed, designed by the Federal Reserve System (referred as the Fed) of United States of America (USA), which simulates inflation rates for a given interbank interest rate over a period of time. The aim is to find out an interbank interest rate that leads to inflation rates closest to the target.

Though the computer models are cheaper / feasible alternatives of the unobservable / expensive physical processes, realistic simulators of complex physical phenomena can also be computationally demanding, i.e., one run may take from seconds/minutes to days/months. In such a scenario, a statistical metamodel or surrogate is often used to emulate the outputs of the simulator and draw inference based on the emulated surrogate. In computer experiment literature, Gaussian process (GP) model is perhaps the most popular class of statistical surrogate because of its flexibility, closed form predictors, and ability to incorporate various uncertainties in model specification (Sacks et al. (1989), Santner et al. (2003)).

Given that the simulator is expensive (computationally or otherwise), one has to be very careful in selecting the input points while training the surrogate. One efficient method is to use a sequential design framework that exploits the overall objective. For instance, Jones et al. (1998) developed an expected improvement (EI)-based design scheme for estimating global minimum, and in the same spirit Ranjan et al. (2008) proposed another EI criterion for estimating a pre-specified contour. See Bingham et al. (2014) for a brief review. Since the simulator under consideration returns time-series response, the sequential approach with standard GP model by Ranjan et al. (2008) cannot directly be used.

We propose scalarizing this \emph{functional inverse problem} by first computing the Euclidean distance, $w(x) = \|g(x)-g_0\|$ for every $x$ and then find the global minimum of $d(\cdot)$ using the EI-based sequential approach with GP model proposed by Jones et al. (1998). Examples in Section~4 illustrates that among all realizations of GP that emulate $w(\cdot)$, a few (in fact numerous) realizations would give negative $\hat{w}(x)$ for $x$ near the global minimum, which is unacceptable as $w(\cdot)$ is the Euclidean distance. To alleviate this theoretical glitch, we first propose building a non-stationary surrogate of $y(x) = \log(w(x))$ via Bayesian Additive Regression Tree (BART) model (Chipman et al. 2010) and then find the global minimum of $y(\cdot)$ using the EI-based design scheme as in Chipman et al. (2012).

The rest of the article is organized as follows: Section~2 presents a brief review of the statistical surrogates and the sequential design scheme. Section~3 discusses the functional inverse problem, the scalarization step and the resultant scalar inverse problem. In Section~4, we present the results on the performance comparison of $y$-inverse and $w$-inverse via both test functions and two real applications: calibration of TDB model, and CTF model. Finally Section~5 concludes the paper with a few remarks.\\

\sectionn{Review}

{ \fontfamily{times}\selectfont
 \noindent In this section, we briefly review the GP model, key features of BART model as a non-stationary surrogate for computer models, and the EI-based sequential design scheme for estimating the global minimum of $w$- and $(\log(w) = ) y$- surface. For this section, we assume that the simulator returns scalar response $y(x)$ for $d$-dimensional input $x\in [0,1]^d$.

\subsection{Gaussian process models}

Sacks et al. (1989) suggested using realization of a GP for emulating deterministic computer simulator outputs. Since then several variations have been proposed for building surrogates of expensive computer models (see Santner et al. 2003, Rasmussen and Williams 2006). The simplest version of a GP model with $n$ training points, $(x_i,y_i), i=1,...,n$, is given by
\begin{equation}
y(x_i) = \mu + Z(x_i), \quad i=1,2,...,n,
\end{equation}
where $\mu$ is the mean term and $Z(x)$ is a GP with $E(Z(x))=0$ and spatial covariance structure defined as $Cov(Z(x_i), Z(x_j)) = \Sigma_{ij} = \sigma^2 R(\theta; x_i,x_j)$, denoted by $Z(x)\sim GP(0, \sigma^2R(\theta))$. The most important component of the GP model, which makes it very flexible, is the correlation structure. Gaussian correlation is perhaps the most popular because of its properties like smoothness and usage in other areas like machine learning and geostatistics, whereas, both power-exponential and Matern can be thought of as generalizations of the Gaussian correlation. The power-exponential correlation is given by
\begin{equation}
 R(x_i, x_j) = \exp\left(-\sum_{k=1}^d \theta_k |x_{ik}-x_{jk}|^{p_k}\right),
\end{equation}
where $0< p_k \le 2$ are the smoothness parameters, and $\theta = (\theta_1,...,\theta_d)$ measures the correlation lengths. Gaussian correlation corresponds to $p_k=2$ for all $k=1,2,...,d$. This model can be fitted either via the maximum likelihood estimation (MLE) or a Bayesian approach. Under the likelihood approach, the best linear unbiased predictor of $y(\mathbf{x}^*)$ is 
\begin{equation}
	\label{eqn-gp-estimation}
	\hat{y}(\mathbf{x}^*) = \hat{\mu} + \mathbf{r}(\mathbf{x}^*)^T\mathbf{R}^{-1}(\mathbf{y}-\hat{\mu}\mathbf{1}_n),
\end{equation}
where $\mathbf{r}(\mathbf{x}^*) = [\text{corr}(z(x^*), z(x_1)), \text{corr}(z(x^*), z(x_2)),...,\text{corr}(z(x^*), z(x_n))]$, and the associated uncertainty (mean squared error) is
\begin{equation}
	\label{eqn-gp-estimated-variance}
	s^2(\mathbf{x}^*) = \hat{\sigma}^2\left(1 - \mathbf{r}(\mathbf{x}^*)^T\mathbf{R}^{-1}\mathbf{r}(\mathbf{x}^*)\right).
\end{equation}

It turns out that the actual implementation of both methods (MLE and Bayesian) suffer from numerical instability in computing the determinant and inverse of $R$. The problem of numerical instability is certainly more pronounced for GP models with Gaussian correlation as compared to other power-exponential and Matern correlation. See Ranjan, Haynes and Karsten (2011) for more details. Popular implementations of the GP model like mlegp (Dancik 2008), GPfit (MacDonald et al. 2015), GPmfit (Butler et al. 2014), and DiceKriging (Roustant et al. 2012) use some sort of numerical fix to overcome the computational instability issue. We used GPfit in R for all implementations of the GP model.

If the process is believed to be non-stationary (e.g., $\log(w(x))$ in this case), one possibility is to modify the covariance parameters to capture this variation, for example, $\sigma$ and $\theta$ can be a function of $x$. Another popular alternative is to use ``treed Gaussian process" (TGP) model proposed by Gramacy and Lee (2008), where the main idea is to split the input space into rectangles and fit separate GP model in each rectangle. Chipman et al. (2012) found TGP somewhat unreliable and proposed the usage of a more flexible non-parametric statistical metamodel called BART (Bayesian additive regression tree).

\subsection{BART model}

Chipman et al. (2010) proposed BART for approximating the conditional mean of the response given the data using a sum of regression trees. In our context, the computer simulator output can be emulated using the BART model as
\begin{equation}
	\label{eqn-bart}
	y_i = \left(\sum_{j=1}^m h(x_i;T_j,M_j)\right) + \epsilon_i,
\end{equation}
where $\epsilon_i\sim\mathcal{N}(0, \sigma^2_{\epsilon})$. This model assumes the existence of $m$ binary trees $T_j$ ($j=1,...,m$) each containing a set of interior node decision rules and $b$ terminal nodes. The parameter $M_j = (\mu_{j1},\mu_{j2},...,\mu_{jb})$ represents the set of mean response parameters at each terminal node of $T_j$. The predicted outcomes of the computer simulator are obtained by sequentially following the decision rules for each tree $T_j$ until reaching a terminal node, and then summing up these terminal node values (i.e., $\mu_{ju}$, for $j\in(1,...,m)$, $u\in(1,...,b)$). Thus, viewed as a function of $x$, tree model $h(x; T_j, M_j)$ produces a piecewise-constant output. 

The ``ensemble" of $m$ such tree models in (\ref{eqn-bart}) makes the BART model very flexible. It is capable of incorporating higher-dimensional interactions, by adaptively choosing the structure and individual rules of the $T_j$'s. Furthermore, many individual trees $(T_j)$ may place split points in the same area, allowing the predicted function to change rapidly nearby, effectively capturing non-stationary behaviour such as abrupt changes in the response.

The model fitting is done via a Markov Chain Monte Carlo (MCMC) \textit{Bayesian backfitting algorithm}. Each iteration of the this algorithm generates one draw from the posterior distribution of $y_i=\sum_{j=1}^m h(x_i;T_j,M_j)$. Let $y^*_{1},...,y^*_{N}$ denote $N$ draws of the posterior of $y(x^*)$. Given a reasonable burn-in period $B$ and thinning constant $\tau$, estimates of the computer simulator can be obtained as simply the average of the observed posterior draws, i.e.,
\begin{align}
	\hat{y}(x^*) = \frac{1}{K}\sum^K_{k=1} y^*_{k},
\end{align}
where $K=(N-B)/\tau$. Estimates of the predicted variation can similarly be computed as the $5^{th}$ and $95^{th}$ quantiles or standard deviation of observed posterior draws. We follow the same formulation of prior as in Chipman et al. (2010), i.e., i) $T_1, \ldots, T_m$ are i.i.d., ii) all elements of $M_1, \ldots, M_m$ are i.i.d. given all $T$'s, and $\sigma_{\epsilon}$ is independent of all $T$'s and $M$'s. 

For applying BART to our deterministic computer experiment, we relax the default prior, $\mu\sim N(0,\sigma^2_{\mu}) = N(0,1/(4k^2m))$ with $k=2$, to $k=1$. Choosing a smaller value of $k$ increases the prior variance of output applying less shrinkage (or smoothness) of the response. The deterministic assumption also requires modification to the prior on $\sigma_{\epsilon}$.  This is accomplished with the same inverted-chi-squared prior for $\sigma^2_{\epsilon}$ as in Chipman et al. (2010), using their recommended value of 3 degrees of freedom, and anchoring the 90th percentile of the $\sigma_{\epsilon}$ prior at {\tt 0.20$\times$sd(y)}, where {\tt sd(y)} is the sample standard deviation of the training $y$ values.  This strategy facilitates MCMC mixing for BART, and can also be considered as having a nugget in GPs, for numeric stability and predictive accuracy. 

In this paper we use the freely available R package BayesTree for implementing all BART models. Recently, Pratola et al. (2014) have developed a computationally more efficient surrogate model equipped with parallel computing functionality.

\subsection{Sequential design}

Given the fixed budget of $n$ simulator evaluations, a naive method of estimating a pre-specified feature of interest (FOI) would be to first choose $n$ input (training) points in a space-filling manner, build (train) the surrogate, and then estimate the FOI from this fitted emulator. Popular space-filling designs in computer experiment applications are Latin hypercube designs (LHDs) with space-filling properties like maximin, minimum pairwise coordinate correlation, orthogonal arrays, etc.  

It has been shown via numerous illustrations in the literature that any reasonably designed sequential sampling scheme outperforms the naive one-shot design approach. The key steps of a sequential design framework is summarized as follows:

\begin{enumerate}
	\item Choose an initial set of points $D=(x_1,...,x_{n_0})$ from the input space $[0,1]^d$.
	\item Obtain the vector of corresponding simulator outputs $Y =y(D)$.	\item Fit a statistical surrogate using the data $D$ and response vector $Y=y(D)$.
	\item Estimate the feature of interest (FOI) from the trained (fitted) surrogate. 
	\item If (the budget is exhausted, or a stopping criterion has met), then exit, else continue.	
	\item \begin{enumerate}
			\item Find a new input point (or set of points) $x_{new}$ by optimizing a \textit{merit-based criterion} (e.g., EI criterion).
			\item Obtain $y(x_{new})$ from the computer simulator.
			\item Append $x_{new}$ and $y(x_{new})$ to the current design and computer simulator response, respectively, forming $D = D\cup \{x_{new}\}$ and $Y=Y\cup \{y(x_{new})\}$.
		\end{enumerate}
	\item Go back to Step~3 (refit the surrogate with the updated data).
\end{enumerate}

For the surrogate in Step~3, we consider both the GP model and BART, and the FOI is the global minimum. The most important part of this sequential framework is Step~6(a). Though one can easily come up with a merit-based criterion, proposing a good one, that can lead to the global minimum in the fewest number of follow-up runs, is not easy. 

Jones et al. (1998) proposed the most popular sequential design criterion in computer experiment literature called the \emph{expected improvement} (EI) with the objective of finding the global minimum of an expensive deterministic computer simulator. The proposed \textit{Improvement} at an untried point $x^*$ is given by $I(x^*)=\max(y_{min}-y(x),0)$, where $y_{min}$ is the current best estimate of the global minimum, and $y(x) \sim N(\hat{y}(x), s^2(x))$. Then, EI is simply the expected value of $I(x^*)$ with respect to the predictive distribution. That is,
\begin{align}
	E[I(x^*)] =& (y_{min}-\hat{y}(x^*))\Phi(u) + s(x^*)\phi(u)~,~~~u=\frac{y_{min}-\hat{y}(x^*)}{s(x^*)},
\end{align}
where $\phi$ and $\Phi$ are the standard normal probability density function and cumulative distribution functions, respectively. Finally, the new point to be added to the experimental design is chosen as the point with the largest measured EI value, that is
\begin{align}
	x_{new} = \arg\max_{x^*}E[I(x^*)].
\end{align}

Inspired by Jones et al. (1998), a host of EI criteria have been proposed for estimating different FOIs. See Bingham, Ranjan and Welch (2014) for a recent review on such merit-based design criteria. Finding the optimal follow-up design point also depends on the accuracy of EI optimization. It turns out that the EI surfaces are typically multi-modal, and the location and number of prominent modes/peaks changes from iteration-to-iteration. Popular optimization techniques used for EI optimization include, genetic algorithm (Ranjan et al. 2008), particle swarm optimization (Butler et al. 2014), multistart newton-based methods (MacDonald et al. 2015), and branch-and-bound algorithms (Franey et al. 2010). Thus, one should be careful in choosing the follow-up points to achieve the optimal improvement.

}

\sectionn{Inverse Problem for Time-series Response}
\label{sec:MLE}

{ \fontfamily{times}\selectfont
 \noindent
In this paper, we assume that the computer simulator is deterministic, takes a $d$-dimensional input $x \in [0,1]^d$ and returns a time-series response $g(x)=\{g(x,t), t=1,2,..., L\}$. Our main objective is find $x \in [0,1]^d$ such that $g(x,t)\approx g_0(t)$ for all $t=1,2,...,L$, where $g_0$ is a pre-specified process output (e.g., the true observed field data in the TDB model application, or the target inflation rates in the CTF model application). This inverse problem with time-series / functional response is also refereed to as the calibration problem, wherein, the main objective is to calibrate the computer model at a certain $x_0$ so that the computer simulator produces desirable (realistic / close to target) response.

The key idea is to propose a scalarization strategy that transforms the functional inverse problem to a minimization problem for a scalar-valued simulator. That is, for every $x\in [0,1]^d$, first transform the simulator output $g(x)=\{g(x,t), t=1,2,...,L\}$ to $w(x)$ as follows:
$$ w(x) = \|g(x)-g_0\| = \sqrt{\frac{1}{L}\sum_{t=1}^L |g(x,t)-g_0(t)|^2},$$
then find the global minimum of $w(x)$. We use EI-based sequential design scheme with GP model as a surrogate (by Jones et al. (1998)) to efficiently minimize $w(x)$.

It is important to note that the predicted realizations of $w(x)$ under the GP model will not always be positive. For instance, the prediction near the global minimum has a good chance of being negative (see Figure~1(b) in Section~4). This may not be critical from the inverse problem's viewpoint, as the negative values of $\hat{w}(x)$ near the global minimum in some iterations of the sequential design scheme do not hinder the efficiency in finding the location of the global minimum, i.e., the inverse solution for $g(\cdot)$. However, a negative value of $\hat{w}(x)$ has no real / feasible inverse mapping to $g(x)$. 

One possibility is to fit a log-GP model, i.e., fit a GP model to $\log(w(x))$. However, as shown in Figures~1(c) and 2(c), $\log(w(x))$ is often non-stationary near the global minima, and a standard GP would not be suitable for this either. We use the flexible BART model (Chipman et al. (2010)) for emulating $y(x) = \log(w(x))$. Subsequently, we use the EI-based sequential design scheme with the BART-based surrogate for efficient minimization of the scalarized response $y(x)$ (e.g., in Chipman et al. (2012)). 

Next we use simulated and real-life examples to compare the performance of the EI-BART method for minimizing $y(x)$ with the naive EI-GP method for minimizing $w(x)$.
 }

\sectionn{Results}
\label{sec:K-M}

{ \fontfamily{times}\selectfont
 \noindent
 In this section, we consider the functional inverse problems for one test function with slight variations that enables $1$-, $2$- and $3$- (dimensional) inputs, TDB model with six-dimensional inputs, and CTF model with 1-dimensional input. For all examples, we use both EI-BART and EI-GP approaches with the same sequential settings, i.e., same $n_0$ (initial design size) + $n_{new}$ (follow-up points). In fact, the initial design points are same for both methods. Moreover, we keep the GP and BART settings same for all examples.\\

\noindent{\bf Example 1.} Suppose the simulator $g(x,t)$ takes an input $x\in [0,1]$ and generates a time-series response with $t=0.5, 0.52, ..., 2.48, 2.50$, as per the following model:
\begin{equation}
g(x,t) = \frac{\sin(10 \pi t)}{2t} + |t-1|^{(2+4x)}.
\end{equation}
Let the true field data correspond to $x_0=0.5$. Figure~\ref{fig:testfn_1d} shows the true field data (solid red curve) and the computer model outputs for a few randomly generated inputs $x\in[0,1]$ (shown in blue dotted curves).

\begin{figure}[h!]\centering
\includegraphics[width=5in]{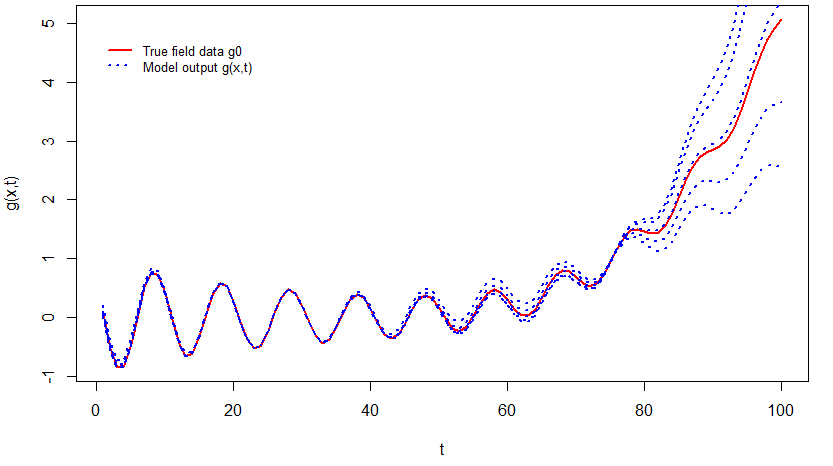}
\caption{A few computer model outputs and the true field data for Example~1.}\label{fig:testfn_1d} 
\end{figure}

Figure~\ref{fig:1d_seq_GP} illustrates results for EI-GP implementation with $n_0=5$ and $n_{new}=10$. The estimated $x_{opt}$ is $0.5000$ and the corresponding minimum $w(x_{opt})$ is $0.0004$.\\

\begin{figure}[h!]\centering
\subfigure[Based on initial design]{\includegraphics[width=3.15in]{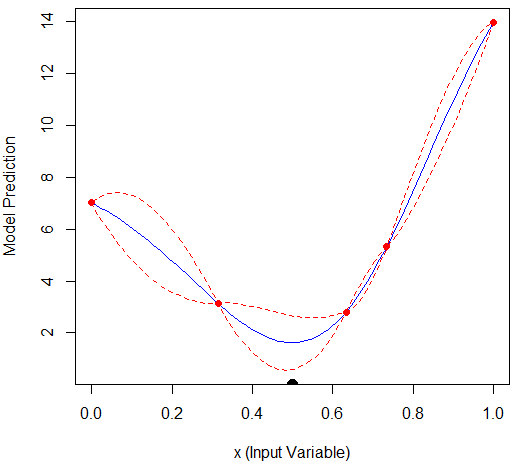}}
\subfigure[After adding 1 follow-up point]{\includegraphics[width=3.15in]{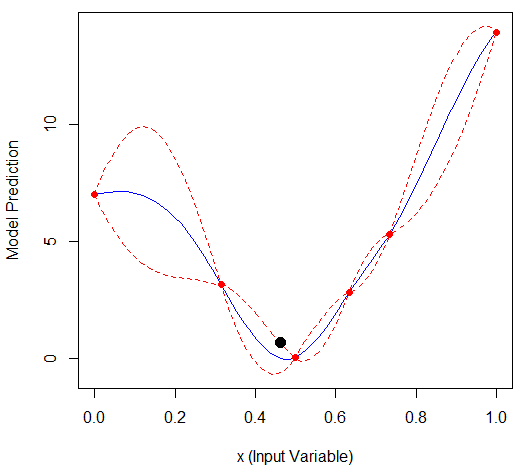}}
\subfigure[After adding 15 follow-up points]{\includegraphics[width=3.15in]{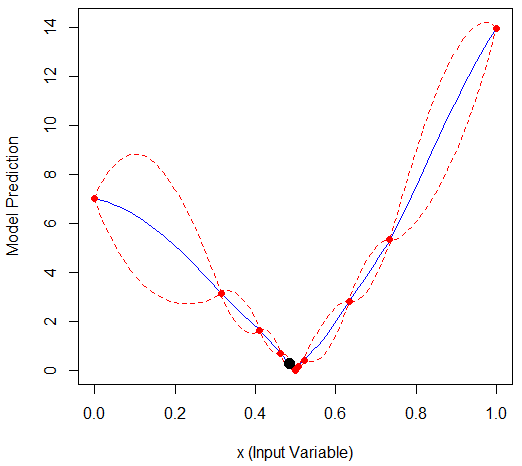}}
\subfigure[Running estimate of $f_{min}$ values]{\includegraphics[width=3.15in]{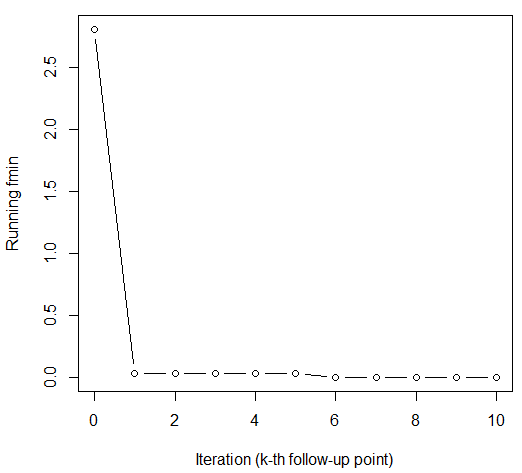}}
\caption{Sequential optimization of $w(x)$ via the GP-based emulator for Example~1.}\label{fig:1d_seq_GP} 
\end{figure}

It is clear from Figure~\ref{fig:1d_seq_GP}(d) that the global minimum was located very quickly in this sequential procedure, which was expected given the simplicity of the test function. Figure~\ref{fig:1d_seq_GP}(b) also shows that many realizations of the GP model that emulate the training data would give negative value of $w(x)$ in the vicinity of the global minimum (the confidence band shown in red dashed curves around the blue solid curve goes below zero in the interval $[0.4, 0.6]$).

Figure~\ref{fig:1d_seq_bart} shows EI-BART illustration with the same sequential settings as in Figure~\ref{fig:1d_seq_GP} for finding the minimum of $y(x)=\log(w(x))$. The estimated $x_{opt}$ is $0.4999$ and the value of the corresponding minimum $y(x_{opt})$ is $-6.6723$ (with $w(x_{opt}) = 0.0012$).

\begin{figure}[h!]\centering
\subfigure[Based on initial design]{\includegraphics[width=3.15in]{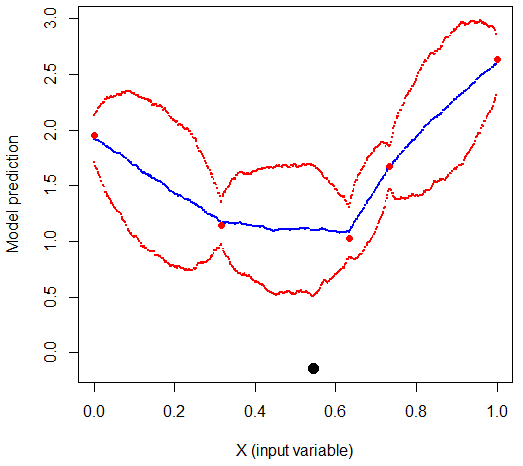}}
\subfigure[After adding 1 follow-up point]{\includegraphics[width=3.15in]{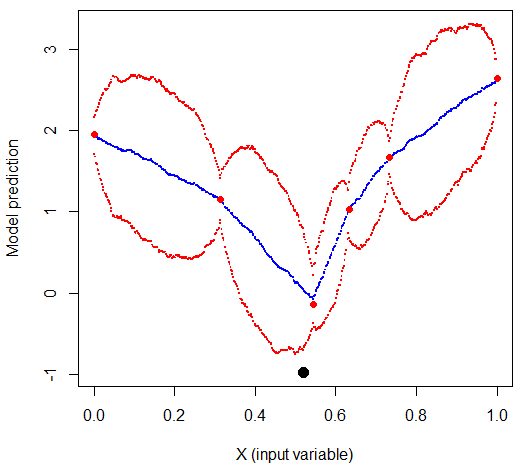}}
\subfigure[After adding 15 follow-up points]{\includegraphics[width=3.15in]{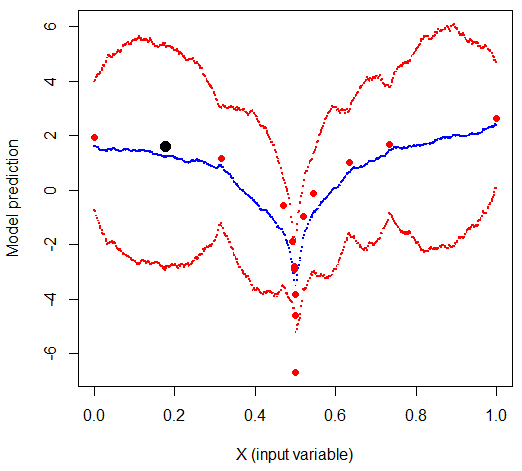}}
\subfigure[Running estimate of $f_{min}$ values]{\includegraphics[width=3.15in]{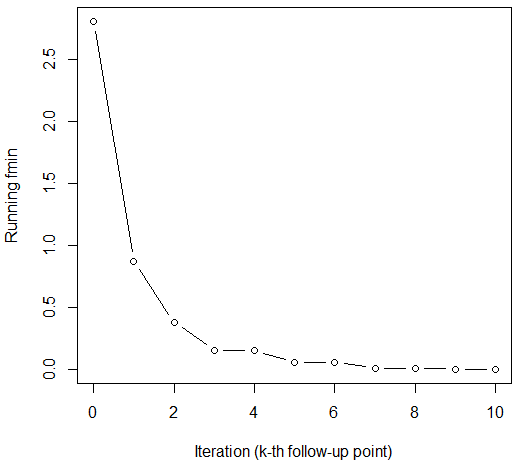}}
\caption{Sequential optimization of $\log_e(w(x))$ via the BART-based emulator for Example~1.}\label{fig:1d_seq_bart} 
\end{figure}

Both EI-BART and EI-GP find the global minimum, but EI-GP exhibit much faster convergence. This is also expected as BART is typically a bit more data-hungry than the GP models. Further note from Figure~\ref{fig:1d_seq_bart}(c) that $\log(w(x))$ surface is highly non-stationary near the global minimum. \\

\noindent{\bf Example 2.} Consider the same test function as in Example~1, with a small modification that would allow two-dimensional inputs $x=(x_1,x_2)\in [0,1]^2$ and generates time-series response in the same time domain. That is, 
\begin{equation}
g(x,t) = \frac{\sin(10 \pi t)}{(1+2x_1) t} + |t-1|^{(2+4x_2)}.
\end{equation}
Let the true field data correspond to $x_0=(0.5,0.5)$. Figure~\ref{fig:testfn_2d} shows the true field data (solid red curve) and a few simulator outputs (blue dotted curves).

\begin{figure}[h!]\centering
\includegraphics[width=6in]{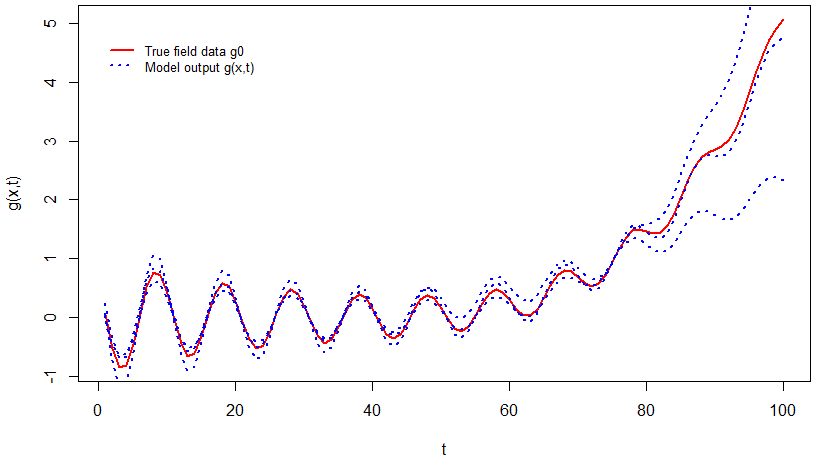}
\caption{A few model outputs and the true field data for the simulator in Example~2.}\label{fig:testfn_2d} 
\end{figure}

For this inverse problem, we used $n_0=10$ initial design points and $n_{new}=20$ follow-up points as per the EI criterion. Figure~\ref{fig:2d_seq_GP} shows the illustration of the EI-GP approach. The estimated $x_{opt}$ is $(0.4932, 0.4985)$ and the corresponding minimum $w(x_{opt})$ is $0.0344$.

\begin{figure}[h!]\centering
\subfigure[Based on initial design]{\includegraphics[width=3.15in]{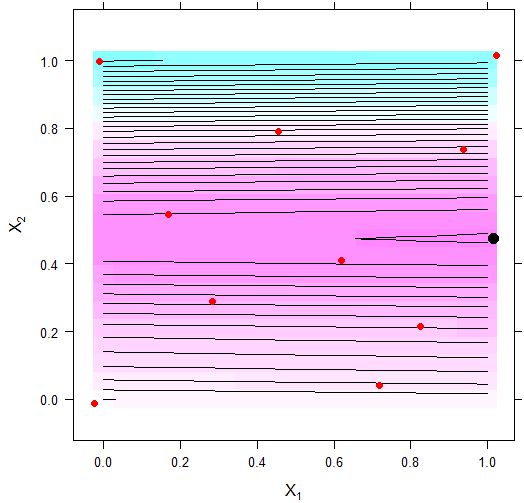}}
\subfigure[After adding 1 follow-up point]{\includegraphics[width=3.15in]{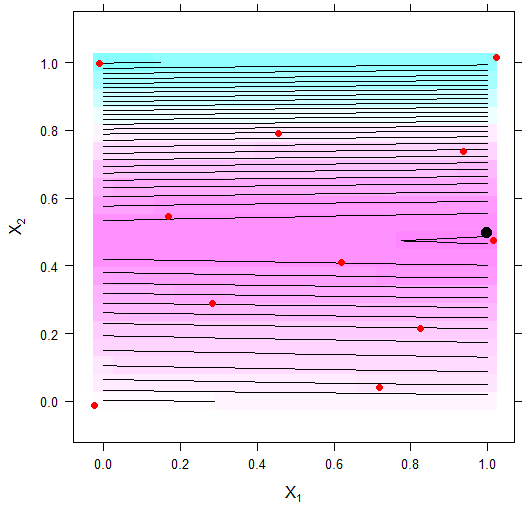}}
\subfigure[After adding 20 follow-up points]{\includegraphics[width=3.15in]{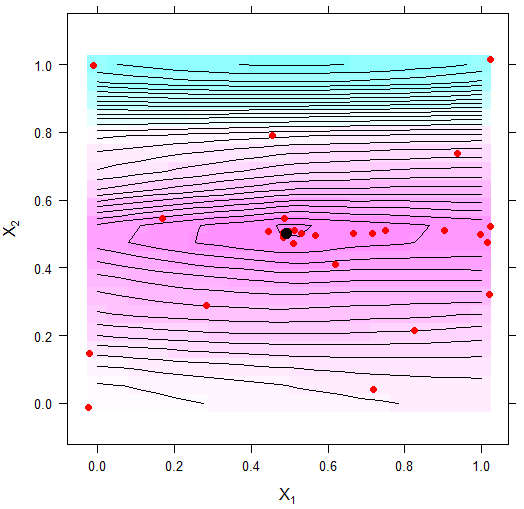}}
\subfigure[Running estimate of $f_{min}$ values]{\includegraphics[width=3.15in]{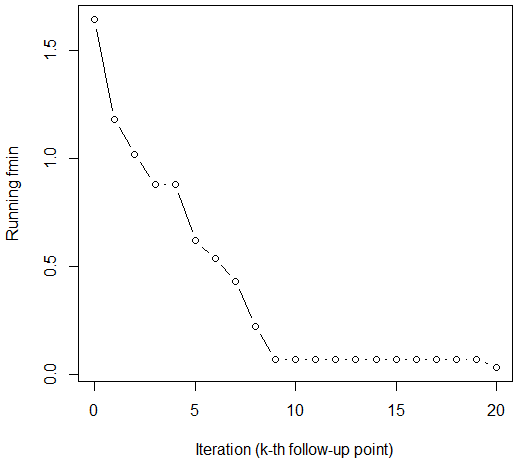}}
\caption{Sequential optimization of $w(x)$ via the GP-based emulator for Example~2.}\label{fig:2d_seq_GP} 
\end{figure}

Figure~\ref{fig:2d_seq_GP}(d) shows that a decent value of the global minimum has been found after $9-10$ follow-up trials. Of course, the efficiency can perhaps be improved by exploring different $n_0$ and $n_{new}$ combination. As expected EI-BART requires a few more points to attain the same accuracy level (see Figure~\ref{fig:2d_seq_bart}).
The estimated $x_{opt}$ is $(0.4897, 0.4995)$ and the corresponding minimum $y(x_{opt})$ is $-3.39$ (i.e., $w(x_{opt}) = 0.0335$).\\

\begin{figure}[h!]\centering
\subfigure[Based on initial design]{\includegraphics[width=3.15in]{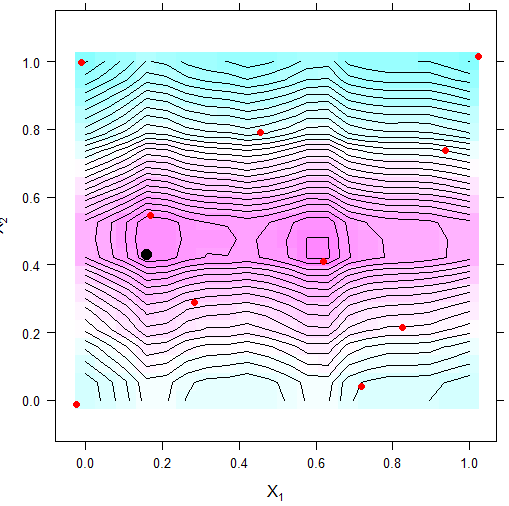}}
\subfigure[After adding 1 follow-up point]{\includegraphics[width=3.15in]{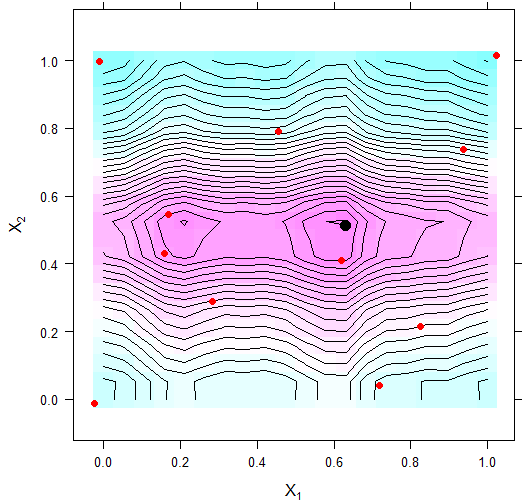}}
\subfigure[After adding 20 follow-up points]{\includegraphics[width=3.15in]{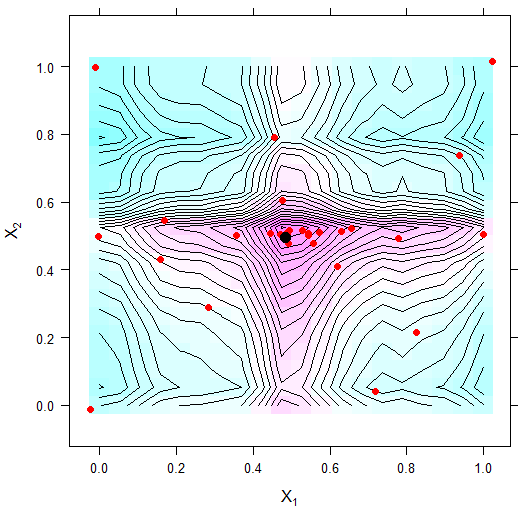}}
\subfigure[Running estimate of $f_{min}$ values]{\includegraphics[width=3.15in]{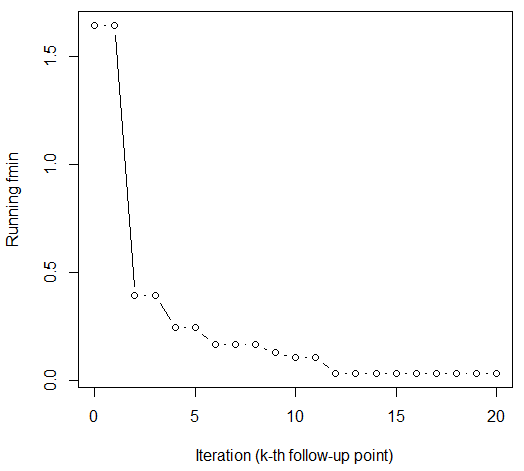}}
\caption{Sequential optimization of $y(x)$ via the BART-based emulator for Example~2.}\label{fig:2d_seq_bart} 
\end{figure}

\noindent{\bf Example 3.} Again we consider the same base example (as in Example~1) with a slight twist to the simulator to allow a three-dimensional input $x=(x_1,x_2,x_3)\in [0,1]^3$, i.e.,
\begin{equation}
g(x,t) = \frac{\sin\left(10 \pi t^{(2x_3)}\right)}{(1+2x_2) t} + |t-1|^{(2+4x_3)}.
\end{equation}
Furthermore, we assume that the true field data correspond to $x_0=(0.5,0.5,0.5)$. Figure~\ref{fig:testfn_3d} shows the true field data (solid red curve) and a few simulator outputs (blue dotted curves). This inverse problem appears to be a little more challenging than the previous ones.

\begin{figure}[h!]\centering
\includegraphics[width=6in,height=3.5in]{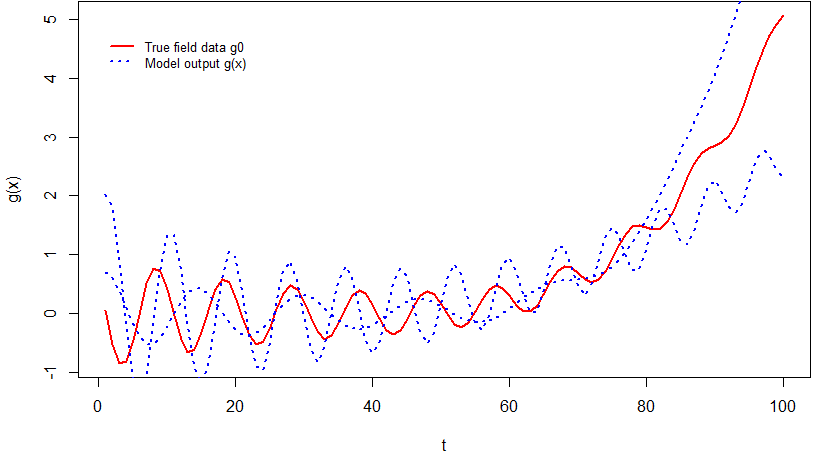}
\caption{A few computer model outputs and the true data for Example~3.}\label{fig:testfn_3d} 
\end{figure}

As the input dimension grows, we have increased the initial design size and the overall budget to $n_0=20$ and $n_{new}=30$ respectively. Figure~\ref{fig:3d_seq_running} compares the performance of EI-GP and EI-BART. As expected EI-GP is leading by a small margin, but EI-BART is a theoretically more correct methodology to follow.

\begin{figure}[h!]\centering
\subfigure[EI-GP]{\includegraphics[width=3.15in]{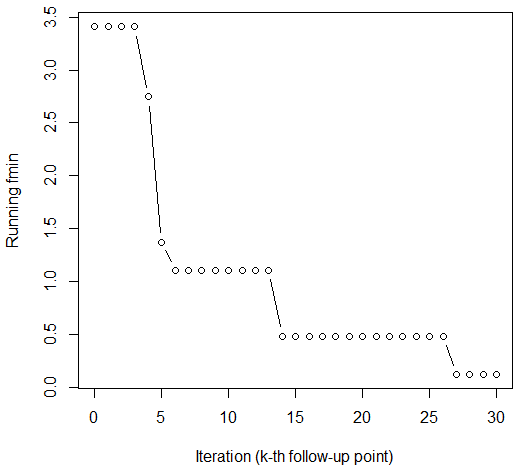}}
\subfigure[EI-BART]{\includegraphics[width=3.15in]{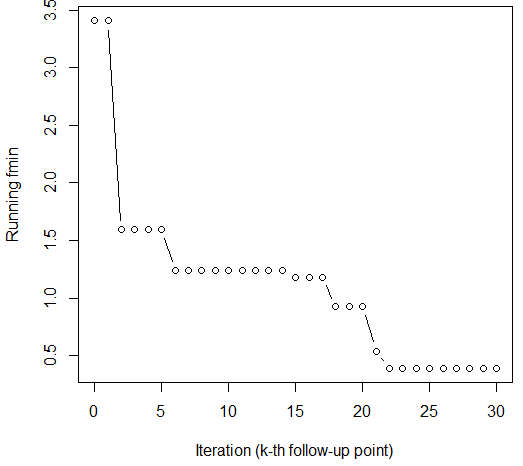}}
\caption{Running estimate of $f_{min}$ obtained under the two methods of finding inverse solution for Example~3.}\label{fig:3d_seq_running} 
\end{figure}

Under EI-GP, the final estimate of $x_{opt}$ is $(0.4827, 0.4979, 0.4991)$ and the corresponding estimate of the global minimum of $w(x_{opt})$ is $0.1259$, whereas for EI-BART, the estimated $x_{opt}$ is $(0.5606, 0.4919, 0.5024)$ and the estimated global minimum $d(x_{opt})$ is $0.3913$. Though the final inverse solution ($x$-values) and the estimated global minimum appear to be slightly different for the two methods, the simulator responses at the two inverse solution are almost indistinguishable (see Figure~\ref{fig:3d_seq_best}). \\

\begin{figure}[h!]\centering
\subfigure[EI-GP]{\includegraphics[width=6.0in,height=3.65in]{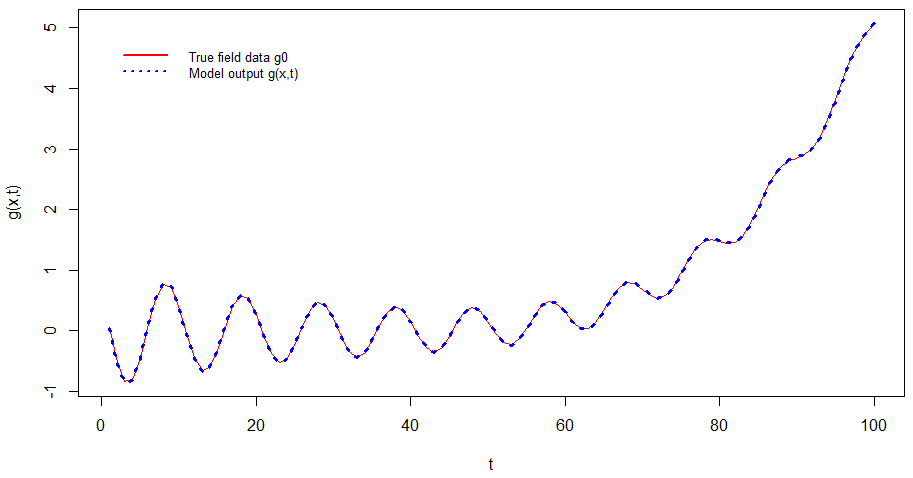}}
\subfigure[EI-BART]{\includegraphics[width=6.0in,height=3.65in]{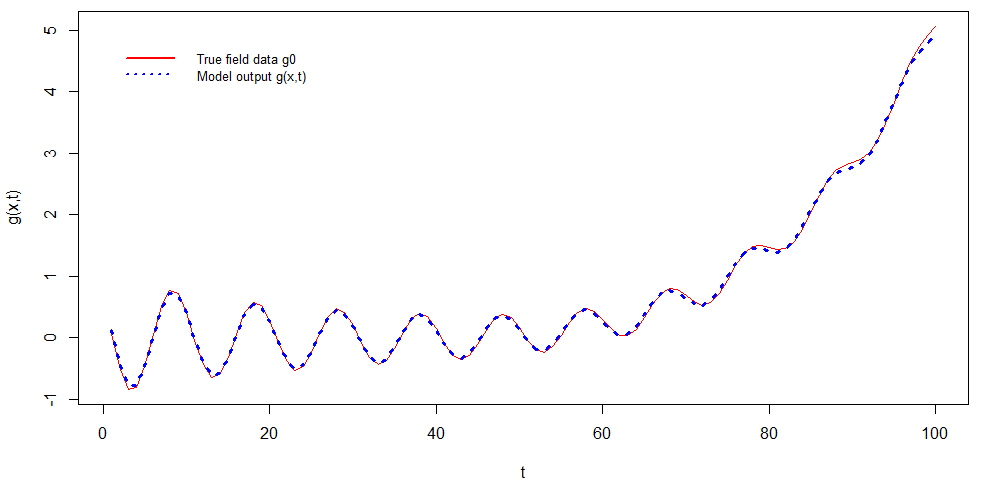}}
\caption{Comparison of the best solution found under EI-GP and EI-BART for Example~3.}\label{fig:3d_seq_best} 
\end{figure}

\noindent{\bf Example 4.} Annapolis Valley in Nova Scotia, Canada is popular for its apple farming.  Unfortunately, apple orchards are susceptible to the infestation of pests. Of particular interest is the Panonychus ulmi (Koch) or European red mite (ERM).

Growth cycle of a mite consists of three stages (1) egg (2) juvenile and (3) adult. These mites start their lives as eggs that are laid in the late summer months of the previous year. Once the temperature rises to a sufficient level the following spring, these eggs hatch and emerge as larvae which further grow to juveniles followed by egg-laying adults. During the summer, adult female ERM lay eggs that hatch during the same season due to the warmer climate. Finally, in mid-to-late August, ERM lay eggs and the cycle repeats itself. 

For deeper understanding of the dynamics of ERM population growth, data collection and analysis is important, but the field data collection from apple orchards is very expensive, as the field experts would have to physically go to the orchard on multiple occasions and count the number of mites (in different stages) from the leves of trees. Tiesmann et al. (2009) proposed a two-delay blowfly (TDB) model that tries to mimic the population growth of these mites. Though there are several parameters of this model, the following six parameters turned out to be very influential:

\begin{itemize}
\item $\mu_4$ - adult death rate
\item $\beta$ - maximum fecundity (eggs laid per day)
\item $\nu$ non-linear crowding parameter
\item $\tau_1$ - (first delay) hatching time of summer eggs
\item $\tau_2$ - (second delay) time to maturation of recently hatched eggs
\item Season - average number of days on which adults switch to laying winter eggs
\end{itemize}

Though the TDB model returns the population growth of all three stages of mite, we focus only on the growth cycle of ``juveniles'' in this paper. A feasible range of $x$ was elicited by the experts for running the TDB model. Figure~\ref{fig:mite_field_data} presents a few TDB models output along with the corresponding field data collected with significant effort. The main objective of this study is to use the proposed sequential strategy to efficiently calibrate the TDB model so that it returns realistic outputs. That is, find $x$ (six-dimensional)
such that $g(x,t)$ matches (or approximates) the reality.

\begin{figure}[h!]\centering
\includegraphics[width=6in,height=4in]{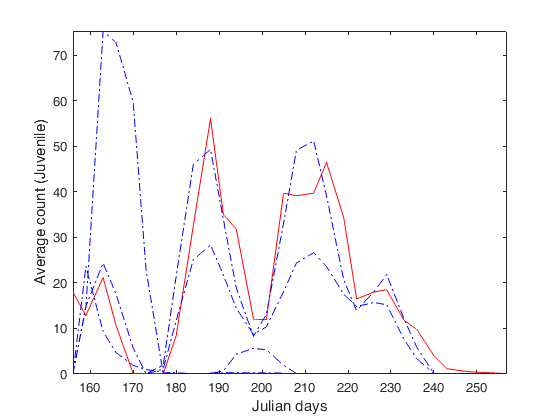}
\caption{A few realisation of juvenile growth curves from the TDB model, and the true average field data collected from the Annapolis valley, NS, Canada.}\label{fig:mite_field_data} 
\end{figure}

As in the earlier examples, we apply both EI-BART and EI-GP with $n_0=60$ and $n_{new}=90$ to find the desired inverse solution. Figure~\ref{fig:TDB_seq_running} shows the running estimate of the global minimum of $w(x)=\|g(x)-g_0\|$.

\begin{figure}[h!]\centering
\subfigure[EI-GP]{\includegraphics[width=3.15in,height=3.2in]{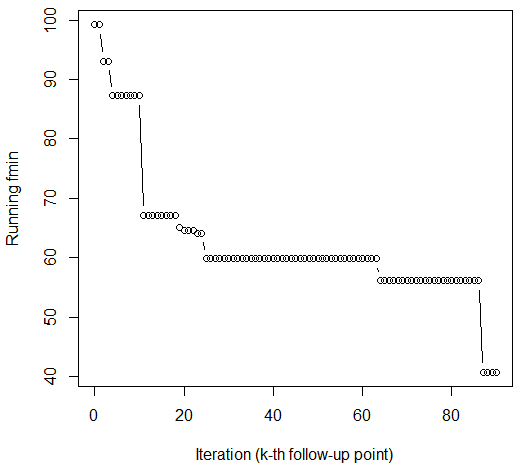}}
\subfigure[EI-BART]{\includegraphics[width=3.15in,height=3.2in]{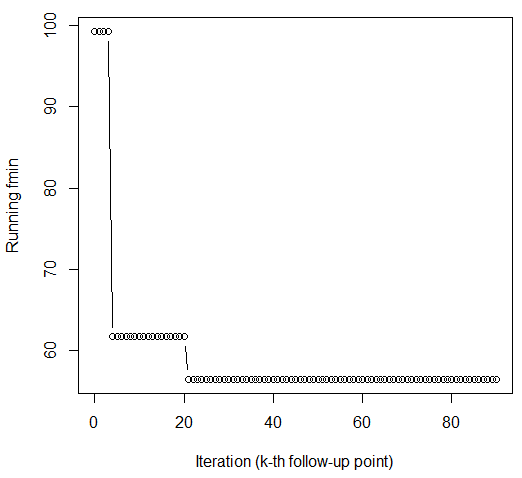}}\caption{Sequential optimization of $w(x)$ for the TDB model outputs.}\label{fig:TDB_seq_running}
\end{figure}

From Figure~\ref{fig:TDB_seq_running} it is clear than EI-GP outperforms EI-BART. ($w(x_{opt})=56.53$ and $40.72$ for EI-BART and EI-GP respectively). Moreover, the best TDB model match obtained via the two methods (Figure~\ref{fig:TDB_seq_bestsol}) show that either additional follow-up points or a different ($n_0, n_{new}$) combination may be required to achieve higher accuracy.\\

\begin{figure}[h!]\centering
\subfigure[EI-GP]{\includegraphics[width=5.75in,height=3.65in]{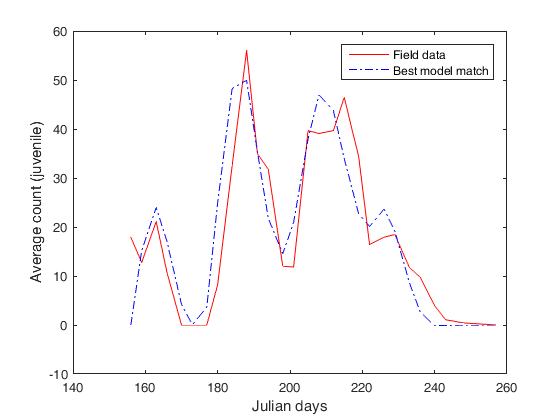}} \\ \subfigure[EI-BART]{\includegraphics[width=5.75in,height=3.65in]{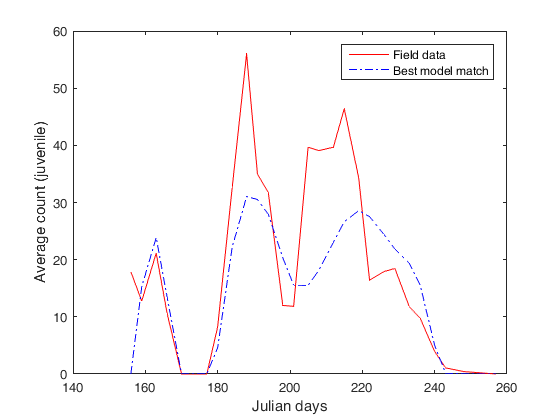}} \caption{Best TDB model runs obtained via the two sequential procedure.}\label{fig:TDB_seq_bestsol} 
\end{figure}


\noindent{\bf Example 5.} Inflation and unemployment are two key tools to measure the financial health of a country. Typically central bank of a country, like the Federal Reserve System (referred to as the Fed) in the United States of America (USA), is mandated to minimize unemployment rate and stabilize prices of goods and services. Central banks aim to do so by controlling the interbank borrowing rate, i.e. the rate of interest at which banks and credit institutions can raise fund overnight from other similar institutions. Decision on interest rate is thus a crucial component of monitory policy for a country.

The monetary policy making body of the Fed, Federal Open Market Committee (FOMC), announces projected inflation (measured using personal consumption expenditures) and unemployment rates based on the analysis of its members for the current year as well as next two years.  In this paper, we focus on the \emph{inflation rates}. Figure~\ref{fig:Fed_model_output} presents the projected inflation rate from the January meeting announcements of FOMC for each year during 2006-2015 (see \url{https://www.federalreserve.gov/monetarypolicy/fomccalendars.htm} and \url{https://www.federalreserve.gov/monetarypolicy/fomc_historical.htm}).


Several interesting theories and models have been proposed thus far to understand the rates projected by the Fed (e.g., Svensson 1997, Huang and Liu 2005). We focus on the simulator  called \emph{Chair-the-Fed} (CTF) (\url{http://sffed-education.org/chairthefed/WebGamePlay.html}), wherein one can select the interbank borrowing rate (i.e., funds rate, the input - $x$) and observe the simulated inflation rates for the next 10 time points (years). The CTF model allows $x$ to vary between $0$ and $20$ with an increment of $0.25$. Figure~\ref{fig:Fed_model_output} depicts a few simulated model runs overlay with the projected rates. Assuming that the Fed funds rate remains static for the next 10 time points, one can play the game (i.e., run the CTF model) and generate a set of inflation fund rates. Our main objective is to find the fund rate ($x$) which generates the inflation rates curve closest to the target (projected values) set by FOMC.

\begin{figure}[h!]\centering
\includegraphics[width=5in,height=3.75in]{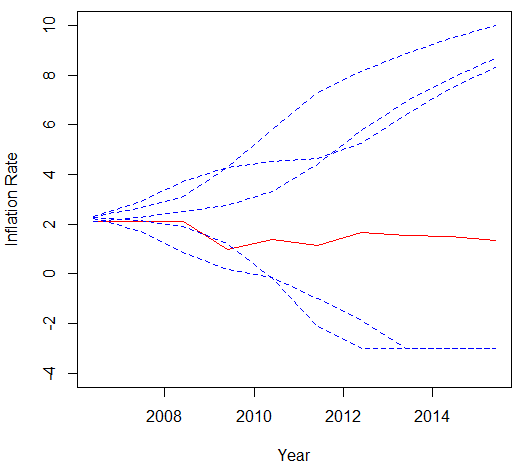}
\caption{A few realisation of the inflation rates curve from the CTF model (blue dashed curves), and the true projected rates set by the Fed and FOMC (red solid curve).}\label{fig:Fed_model_output} 
\end{figure}

Since the CTF simulator is only one-dimensional, we started the sequential approach in both EI-GP and EI-BART with only $n_0=5$ initial points, however, added upto 35 follow-up points to ensure the global minimum. The results from the sequential search of the inverse solution are displayed in Figures~\ref{fig:CTF_seq_running} and \ref{fig:CTF_seq_bestsol}. As shown in Figure~\ref{fig:CTF_seq_running}, EI-BART converged to the global minimum with relatively fewer additional model evaluations as compared to the number points needed by EI-GP.

\begin{figure}[h!]\centering
\subfigure[EI-GP]{\includegraphics[width=3.15in, height=3.25in]{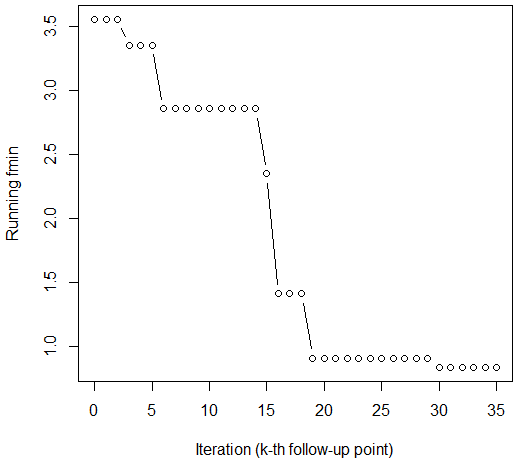}}
\subfigure[EI-BART]{\includegraphics[width=3.15in,height=3.25in]{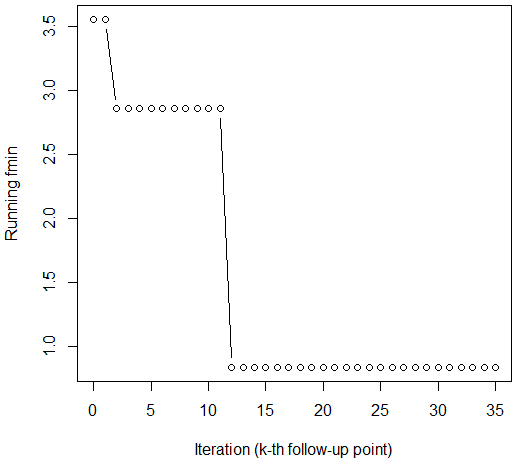}}\caption{Sequential optimization of $w(x)$ for the Chair-The-Fed (CTF) simulator outputs.}\label{fig:CTF_seq_running}
\end{figure}
 
\begin{figure}[h!]\centering
\includegraphics[width=5in,height=3.5in]{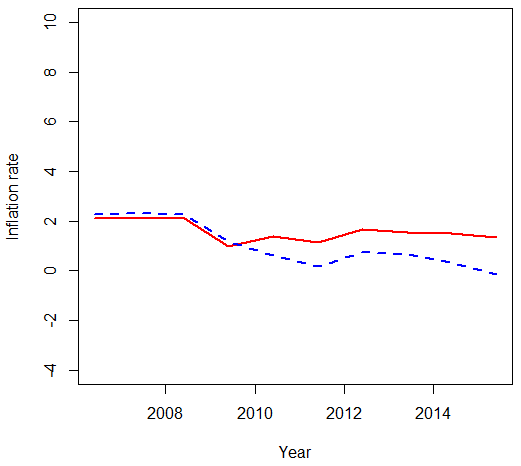} \caption{Best Chair-The-Fed (CTF) model runs obtained via the two sequential procedure.}\label{fig:CTF_seq_bestsol} 
\end{figure}

The discrepancy in the target inflation rates curve and the best match produced by CTF model is perhaps attributed to the discreteness in the $x$-space (CTF allows inputs in the interval $(0, 20)$ with a jump of $0.25$), or the fact that $x$ is not allowed to vary with time, or perhaps additional input variables have to be included to get a closer match. Furthermore, higher efficiency can perhaps be achieved by exploring the right $(n_0, n_{new})$ combination. \\

}

\sectionn{Concluding Remarks}

{ \fontfamily{times}\selectfont
 \noindent 
In this paper, we focused on an inverse problem for deterministic computer simulator with time-series (or functional) outputs. Our main focus was on reducing the complexity of the problem from time-series response to scalar by scalarization, $w(x) = \|g(x)-g_0\|$, where $g_0$ and $g(x)$ are the target (pre-specified) and simulator (time-series) response, respectively. We were also interested in solving the inverse problem with as few simulator runs as possible. This is particularly useful if the simulator is expensive to evaluate, and/or if the input dimension is large which prohibits thorough exploration of the input space. 

The efficiency (minimizing the number of simulator runs) was achieved by using EI-based sequential design scheme and surrogate-based approach. It is explained in Section~3 that the most popular choice of surrogate in computer experiment (GP model) is theoretically inappropriate, however, as illustrated through multiple examples,  EI-GP approach worked equally well for finding the inverse solution. Even if we ignore this theoretical glitch, there is no clear winner between EI-GP and EI-BART. 

There are several interesting issues that should be investigated further and we wish to work on it in our future research endeavours. A few of them are listed as follows: (1) A more thorough comparison between the two methods (EI-GP and EI-BART) should be conducted. For instance, we should use a variety of test functions, repeat the simulations to average out the effect of initial design choice, find optimal ($n_0, n_{new}$) combination in the sequential procedure, and so on. (2) The scalarization process should be further strengthened by using more informative discrepancy measure as compared to Euclidean distance. (3) Does this scalarization procedure affect the likeliness of finding the inverse solution? That is, can this scalarization approach be used for every inverse problem with functional / time-series outputs without risking the accuracy and efficiency? A thorough comparison with non-scalarization based methods should be conducted. (5) Is it straightforward to view the percentile estimation type problem as a generalization of the inverse problem in this setup as well? \\

}

 {\color{myaqua}

 \vskip 6mm

 \noindent\Large\bf Acknowledgments}

 \vskip 3mm

{ \fontfamily{times}\selectfont
 \noindent
 Preliminary work was done by Corey Hodder during his BSc honours thesis at Acadia University, NS, Canada.}

 {\color{myaqua}

}}


\begin{thebibliography}{10}

{\color{black}

\bibitem{} Bingham, D., Ranjan, P., and Welch, W. (2014) ``Sequential design of computer experiments for optimization, estimating contours, and related objectives", Chapter 7 in - \emph{Statistics in Action: A Canadian Outlook}. pp 109 -- 124.


\bibitem{} Butler, A., Haynes, R., Humphries, T.D., and Ranjan, P. (2014) ``Efficient Optimization of the Likelihood Function in Gaussian Process Modeling", \emph{Computational Statistics and Data Analysis}, 73, 40--52.


\bibitem{} Chipman, H., Ranjan, P. and Wang, W. (2012), ``Sequential Design for Computer Experiments with a Flexible Bayesian Additive Model", \emph{Canadian Journal of Statistics}, 40(4), 663--678.

\bibitem{} Chipman, H. A., George, E. I., and McCulloch, R. E. (2010). ``BART: Bayesian additive regression trees." \emph{Annals of Applied Statistics}, 4, 266-298.

\bibitem{} Dancik G.M. and Dorman K.S. (2008).  ``mlegp:  Statistical Analysis for Computer Models of Biological Systems Using R."
\emph{Bioinformatics},24(17), 1966-1967.


\bibitem{} Franey, M., Ranjan, P. and Chipman, H. (2011), ``Branch and Bound Algorithms for Maximizing Expected Improvement Functions" \emph{Journal of Statistical Planning and Inference}, 141, 42 - 55. 

\bibitem{} Gramacy, R. B. and Lee, H. K. H. (2008). ``Bayesian treed Gaussian process models with an application to computer modeling." \emph{Journal of the American Statistical Association}, 103, 1119-1130.

\bibitem{} Huang, K. X., and Liu, Z. (2005). ``Inflation targeting: What inflation rate to target?." \emph{Journal of Monetary Economics}, 52(8), 1435-1462.

\bibitem{} Jones, D., Schonlau, M., and Welch, W. (1998), ``Efficient Global Optimization of Expensive Black-Box Functions," \emph{Journal of Global Optimization}, 13, 455-492.


\bibitem{} MacDonald, K.B., Ranjan, P. and Chipman, H. (2015), ``GPfit: An R package for Fitting a Gaussian Process Model to Deterministic Simulator Outputs" \emph{Journal of Statistical Software}, 64 (12), 1--23.

\bibitem{} Pratola, M.T., Chipman, H.A., Gattiker, J.R., Higdon, D.M., McCulloch, R. and Rust. W.N.  (2014), ``Parallel bayesian additive regression trees." \emph{Journal of Computational
and Graphical Statistics}, 23(3):830-852.

\bibitem{} Ranjan, P., Bingham, D. and Michailidis, G. (2008), ``Sequential Experiment Design for Contour Estimation from Complex Computer Codes", \emph{Technometrics} 50(4), 527-541.

\bibitem{} Ranjan, P., Haynes, R. and Karsten, R. (2011), ``A Computationally Stable Approach to Gaussian Process Interpolation of Deterministic Computer Simulation Data", \emph{Technometrics}, 53, 366-378. 

\bibitem{} Rasmussen, C. E. and Williams, C. K. I. (2006). \emph{Gaussian Processes for Machine Learning}, The MIT Press, Cambridge, MA, USA.


\bibitem{} Roustant, O., Ginsbourger, D. and Deville, Y. (2012). DiceKriging, DiceOptim: Two R Packages for the Analysis of Computer Experiments by Kriging-Based Metamodeling and Optimization. \emph{Journal of statistical software}, 51(1), pp. 1-55.

\bibitem{} Sacks, J., Welch, W. J., Mitchell, T. J., and Wynn, H. P. (1989), ``Design and Analysis of Computer Experiments," \emph{Statistical Science}, 4, 409-423.


\bibitem{} Santner, T. J., Williams, B. J., and Notz, W. I. (2003), ``The Design and Analysis of Computer Experiments," \emph{Springer Verlag, New York.}

\bibitem{} Svensson, L. E. (1997). ``Inflation forecast targeting: Implementing and monitoring inflation targets." \emph{European Economic Review}, 41(6), 1111-1146.

\bibitem{} Teismann H, Karsten R, Hammond R, Hardman J and Franklin J, (2009), ``On the Possibility of  Counter-Productive  Intervention:  The  Population  Mean  for  Blowflies  Models  Can  Be  an Increasing Function of the Death Rate," \emph{J. Biol. Systems},47, 739-757.


}

\end{thebibliography}
\end{document}